# POTENTIAL SPECTRAL MARKERS OF INTRINSIC MAGNETIC FIELDS OF HOT EXOPLANETS IN THE UV RANGE

M. Rumenskikh[1–4], S. S. Sharipov[1], I. F. Shaikhislamov[1–3], A. E. Ivanova[5], M. P. Golubovsky[1,2], K. S. Shcherbakov[1], A. V. Taichenachev[1,3], L. A. Maksimova[6]

1) Institute of Laser Physics, Siberian Branch of the Russian Academy of Sciences, Novosibirsk, Russia
2) Institute of Astronomy, Russian Academy of Sciences, Moscow, Russia
3) Novosibirsk State Technical University, Novosibirsk, Russia
4) Pushkov Institute of Terrestrial Magnetism, Ionosphere and Radio Wave Propagation, Russian Academy of Sciences, Moscow, Russia
5) Space Research Institute, Russian Academy of Sciences, Moscow, Russia
6) Ural Federal University, Ekaterinburg, Russia

email address: marina_rumenskikh@mail.ru

**Abstract**

A method for detecting the intrinsic magnetic fields of exoplanets, based on the effect of atomic alignment—the redistribution of populations among magnetic sublevels under the action of anisotropic radiation from the host star—has been adapted for observations with the Spektr-UV space telescope. A procedure for selecting sensitive transitions is proposed, and triplet absorption lines in the UV range that are sensitive to alignment are identified. On the basis of three-dimensional aeronomic models of the hot Jupiters KELT-20 b and HAT-P-32 b, absorption spectra in these lines are calculated for cases with and without intrinsic magnetic fields of the exoplanets. Using the Spektr-UV exposure-time calculator, the observability of the effect is assessed, and an observing program is proposed, specifying the optimal duration of individual exposures and the number of transits required to detect the signal.

## 1. INTRODUCTION

Exoplanets are relatively new objects of study in astrophysics. Although observations of them began relatively recently compared with those of other astrophysical objects, more than 6400 exoplanets have now been discovered, according to the NASA Exoplanet Archive, and their number continues to grow as new space missions are launched. Unfortunately, Proxima b, the planet in another stellar system closest to Earth, remains beyond the reach of humankind; therefore, all information about extrasolar worlds, including planetary sizes and masses and the composition of their atmospheres, is obtained by optical diagnostics.

One of the most intriguing questions in astrophysics—are we alone in the Universe?—is directly related to exoplanetary atmospheres. The most comprehensive information about them—their composition, structure, and basic physical parameters—is obtained by interpreting observations of spectrally resolved transit absorption with numerical models. The physical, chemical, and kinetic processes included in such a model determine the accuracy with which the physicochemical parameters of planetary systems can be inferred. Thus, unlike non-self-consistent models, self-consistent three-dimensional models that include the denser lower atmospheric layers provide more accurate estimates of planetary mass-loss rates, whereas one-dimensional models do not allow the interaction between planetary and stellar material to be fully accounted for. However, even a complete self-consistent numerical model does not yield a unique solution for the parameters of the absorbing medium from its spectrum, because the atmospheric structure that determines the shape of an absorption line depends on numerous parameters, including the host star's radiation spectrum and the planet's intrinsic magnetic field.

A planet's magnetic field largely determines the structure of its atmosphere and the dynamics of its plasma environment, directly affecting transit absorption profiles. However, direct measurements of exoplanetary magnetic fields are fundamentally inaccessible, forcing researchers to seek indirect diagnostic methods. The most conceptually developed method—the search for electron-cyclotron maser radio emission—has so far yielded no confirmed results: despite optimistic theoretical flux estimates [1–3], observational campaigns with the LOFAR telescope [4] have not led to reliable detections [5,6]. Another method, based on analyzing star–planet interaction (SPI) effects [7], is indirect and suffers from parameter degeneracy when solving the inverse problem of inferring the properties of a medium from its absorption spectrum. The magnetic field is introduced into the model as a free parameter and, together with other parameters—stellar-material mass loss, the intensity of stellar radiation, atmospheric composition, etc.—determines the structure of the modeled atmosphere and the absorption-line profiles. Spectropolarimetric methods are also limited by instrumental and physical factors: according to calculations [8], the fraction of linearly polarized radiation in the He I 1083 nm line is only ~$10^{-4}$ of the absorption-line signal (~1% of the stellar radiation), which is beyond the sensitivity of existing instruments. The lack of independent data with which to test existing methods necessitates the search for fundamentally new optical methods.

In this paper, we propose an alternative method for determining the magnetic fields of exoplanets, based on the quantum-mechanical effect of atomic alignment [9]. Its principal advantage is that it uses absorption lines of planetary atmospheres that have already been observed for several exoplanets. Thus, it is currently possible to determine the intrinsic magnetic fields of specific exoplanets. The idea of applying this effect to astrophysical objects dates back to the work of D. A. Varshalovich [10], who demonstrated the possibility of determining the strength and direction of magnetic fields in cometary tails from the asymmetry of absorption in the sodium doublet induced by anisotropic solar radiation. Atomic alignment was first proposed for exoplanets in [8], where it was shown that the presence of a magnetic field should induce a linear-polarization signal at the level of $10^{-4}$–$10^{-3}$ in the He I 1083 nm line; however, such a weak polarization signal is still inaccessible to modern telescopes, motivating the search for alternative observational manifestations of the effect that are easier to measure. The proposed method is based on a physical mechanism whereby anisotropic radiation from the host star selectively pumps the magnetic sublevels of a lower, long-lived atomic level with total angular momentum $J \geq 1$, producing a nonequilibrium population distribution. After tens of transitions in the system of atoms (or ions), a nonequilibrium distribution also forms among the upper energy sublevels. The population anisotropy (nonequilibrium distribution) arises from the selection rules for electric-dipole transitions: the orbital quantum number $l$ may change only by $\pm 1$, and the total angular momentum $J$ by 0 or $\pm 1$, with the 0–0 transition strictly forbidden. If the process is considered qualitatively, excluding magnetic mixing and mixing between sublevels that give rise to off-diagonal elements of the density matrix, the induction of anisotropy can be described as follows: for some sublevels, there are fewer excitation channels out of them than relaxation channels into them. Populations therefore accumulate in some sublevels, creating a nonequilibrium distribution. In the absence of a magnetic field, alignment causes the relative intensities of the multiplet components to deviate from the equilibrium statistical ratio $2J+1$, which appears as a change in the ratio of absorption-line depths within a triplet. In the presence of even a weak magnetic field, the quantization axis rotates, destroying the induced anisotropy and returning the ratio of absorption-line depths in the triplet to their equilibrium values. Thus, by measuring the ratio of the depths of the triplet components in a transit spectrum, the presence of a magnetic field can be detected directly; the method's sensitivity reaches ~0.001 G. The theoretical basis for detecting spectroscopic signatures of alignment in planetary atmospheres was developed in detail in [9], where the method was first tested against transit observations of He I 1083 nm for two hot Jupiters. Although the calculations in that work were performed for the near-infrared range, the physical principle itself does not depend on wavelength, and several promising transitions of heavy elements lie in the ultraviolet (UV) range. It should be noted that the method described in [9] is currently not applicable to strong magnetic fields of ~10 G.

In recent years, most attention has been paid to near-infrared lines because of the technical characteristics of existing telescopes. Over recent decades, spectrally resolved transit absorption has been observed mainly with ground-based instruments equipped with high-resolution spectrographs (CARMENES, NIRSPEC). These observations have refined our knowledge of the atmospheric composition of several hot exoplanets, which has proved particularly rich in ultra-hot Jupiters (T_eff ≥ 2000 K). According to visible and infrared spectroscopy performed with ground-based telescopes, the atmospheres of such planets contain Na I, K I, Li I, Ca II, Mg I, Mn I, Fe I, Fe II, Cr I, and V I [7,11–23], as well as the widespread species H I, O I, and He I [7,19,24,25]. Some heavy ions—Mg II, Sc II, and Fe II—have been detected in the near-UV range (300–400 nm) using the UVES spectrograph on the VLT [26–28]; however, studies in the far-UV region (91.2–300 nm) require observations from above the Earth's atmosphere.

The first spectrally resolved observations of transit absorption were obtained with the Hubble Space Telescope, operating over the wavelength range 115–2500 nm. These observations revealed several important effects occurring in the upper atmospheric layers of exoplanets and provided valuable information about the composition and structure of hot exoplanets. For example, it was shown that some hot exoplanets possess extended, predominantly hydrogen envelopes that flow beyond their Roche lobes. Thus, observations of the warm Neptune GJ 436 b revealed absorption in the Lyα line: absorption of 60% was detected at velocities of [−120, −50] km $s^{-1}$ [29], corresponding to planetary material moving at these velocities from the star toward the observer. It was also shown that planetary atmospheres contain atoms and highly charged ions of heavy elements [30–33] and that, under certain conditions, energetic neutral atoms (ENAs) form at the interaction boundary between the planetary and stellar winds, increasing absorption in the blue part of the hydrogen-line spectrum [34]. Unfortunately, degradation of the Hubble Space Telescope's UV detectors substantially reduced its sensitivity and prevented the existing information about exoplanetary atmospheres from being refined and expanded; hence the planned launch of the Spektr-UV telescope is expected to be the most significant forthcoming event of the next decade.

In this paper, we propose a systematic approach to identifying UV spectral lines that are sensitive to atomic alignment. Based on the constraints associated with the manifestations of the alignment effect (see the section "Model Description"), we propose a list of UV lines that can indicate the presence or absence of exoplanetary magnetic fields. HAT-P-32 b and KELT-20 b are proposed as suitable observational targets because their host stars emit more strongly in the UV than the corresponding stars of other exoplanets observed through transit absorption [35]. The results, presented as a table of lines selected for observation and calculated transit spectra with and without magnetic fields, will facilitate the detection of exoplanetary magnetic fields in future observations and, in the longer term, the development of planetary magnetism theory. We have also developed specific recommendations for an observing program. The relevant subsection of "Results" provides the rationale for particular observing modes, the observing conditions, and the dependence of the signal-to-noise ratio on exposure time, as obtained using the Spektr-UV exposure-time calculator [36]. In the conclusion, we summarize the recommendations based on our model predictions.

## 2. METHODS

### 2.1 Aeronomy of the upper atmospheres of hot exoplanets

The upper atmospheres of hot exoplanets are calculated using the global three-dimensional multi-fluid magnetohydrodynamic (MHD) model Exo3D, which has previously been used repeatedly to interpret transit observations of hot exoplanets [37–41]. It is based on a multi-fluid approach in which the upper atmospheric layers are modeled by jointly solving the continuity, momentum-conservation, and energy-conservation equations for each fluid. The MHD equations are solved numerically on a nonuniform spherical grid. Time integration is performed using a predictor–corrector method, while a second-order accurate finite-difference scheme is used for the spatial coordinates. Each chemical element and its ions are treated as separate fluids, each with its own density, velocity, and temperature, and their interactions are determined by a combination of chemical, radiative, recombination, and collisional processes. The modeled atmosphere includes neutral and ionized hydrogen and helium (H, $H^+$, He, $He^+$, $He^{2+}$), as well as heavy elements (C, O, Mg, Fe) up to ionization state +2. The ability to vary the relative abundances of heavy elements within the numerical MHD model while calculating spectral lines makes it possible to estimate planetary atmospheric composition and the influence of metallicity on atmospheric structure. We note that in this work the calculations were performed without an intrinsic planetary magnetic field; its effect on alignment was taken into account when post-processing the results in the quantum-mechanical density-matrix calculation: a magnetic field was chosen for which the relative change in sublevel populations attained a steady-state value at the given photoexcitation rate.

Particular attention in the model is paid to a self-consistent description of the level kinetics of heavy elements, calculated simultaneously with the MHD equations at every time step. For oxygen, carbon, and magnesium, a simplified multilevel scheme (Single Level Population Model, SLPM) is used, in which the populations of several low-lying excited levels of neutral and singly ionized atoms are calculated. Unlike the full multilevel approach, transitions between excited states are not considered; instead, photoionization, recombination, electron collisions, and radiative transitions from the ground state are treated independently for each level. This approximation substantially reduces the dimensionality of the kinetic problem and the computational cost, while retaining the ability to describe correctly the strongest resonance lines that determine atmospheric cooling and the formation of observed spectra. In addition, the model includes photoionization from both the ground and excited levels, which is particularly important for calculating the local energy balance and electron-density distribution. The use of a limited number of energy levels is motivated by the fact that low-lying excited states make the principal contribution to cooling and spectral-line formation at temperatures characteristic of hot-exoplanet atmospheres. The validity of this simplification was previously checked using detailed kinetic modeling [37], which showed that the populations of the levels considered are mainly determined by the balance between collisional excitation and radiative decay, with contributions from photoionization and recombination, and depend only weakly on transitions between excited states.

The most substantial simplification concerns the description of iron. Owing to the exceptionally complex electronic structure of Fe I and Fe II, calculating the kinetics of numerous excited levels would require solving a very large system of coupled kinetic equations, which is practically difficult to combine with three-dimensional hydrodynamic calculations. Therefore, the level populations of heavy-element atoms and ions are calculated in the local thermodynamic equilibrium (LTE) approximation. We note that the ion density itself is not in LTE and is calculated during the simulation within the SLPM approach. Energy-level populations are calculated after the hydrodynamic iterations during post-processing. For these calculations, we used all the levels from the NIST Levels database [42] and calculated the partition function and population at the corresponding temperature in each computational cell.

### 2.2 Atomic alignment

Atomic alignment is the redistribution of populations among the magnetic sublevels of the same energy level in a system of atoms and ions, arising from their interaction with anisotropic radiation. For atoms and ions in the atmospheres of hot exoplanets, the host star is the source of such directed, unpolarized radiation. To quantify the alignment effect, Hamiltonian equations for a two-level quantum system of particles of one species—atoms or ions—are solved numerically in the weak-pumping-field approximation for unpolarized, directed stellar radiation [9]. By a two-level system we mean quasi-isolated energy levels of atoms or ions corresponding to transitions from a single lower level to the split triplet fine-structure sublevels. By quasi-isolation we mean that collisions involving the selected particle system are not included in the quantum-mechanical model. These are accounted for in the hydrodynamic model; the selected particle system can be related to the actual planetary atmosphere by taking into account a number of conditions ensuring weak interactions between different particle systems in the radiation-absorption region and a relatively high frequency of transitions between the lower and upper sublevels of the quantum system compared with other processes affecting the populations of the selected two-level system.

The result of numerically solving the system of quantum-mechanical equations is a density matrix describing the population of each magnetic sublevel of an individual energy level. The real part of the trace of this density matrix represents the ratio of the populations of the fine-structure energy sublevels. A change in the population ratio (relative to the equilibrium 2J+1 distribution), produced by a directed radiation field or by the combined action of the radiation field and an intrinsic dipole magnetic field acting on the particle system, can be measured observationally as the ratio of absorption depths of lines within each single triplet. Importantly, by a single triplet we do not mean the grouping of the lines under consideration according to their wavelength separation, as is often done in astrophysical publications, but only transitions between one lower energy level and three upper fine-structure sublevels caused by dipole transitions under the influence of the host star's radiation field on a quantum system of particles of the same species.

The fundamental equations, details of their solution, special cases, limitations, and conclusions of atomic alignment theory are described in detail in [9]. Below, we set out the conditions for applicability of the quantum-mechanical model under which alignment can substantially affect the populations of the upper sublevels and, consequently, the ratio of triplet absorption-line depths:

- Directed radiation field: the angular size of the planet is small, $\Theta \sim 0.05 << 1$ (for HAT-P-32 b with $R_p = 1.78\ R_J$ and $a = 0.03427$). This means that any region of the planet is illuminated at a negligibly small angle to the star–planet axis, which corresponds to the quantization axis, so the condition of directional stellar radiation is maintained.
- The angular momentum of the lower level must satisfy $J_g > 1/2$ to produce population anisotropy (see Fig. 1 in [9]). The optical pumping rate from the lower to the upper levels and the reverse radiative decay rate must be much greater (~30 times) than the decay rates of the lower and upper levels through all other processes and channels.
- The particle collision frequency is much smaller than the frequency of resonance transitions between the lower and upper energy levels: $\nu_s$ (~1–$10^3$ Hz) $<<$ $\nu_\gamma$ (~$10^6$ Hz). This condition is satisfied in most regions of exoplanetary atmospheres where, according to the calculations, absorption occurs in atomic and ionic lines, including triplet lines.
- $\gamma S >> \Gamma$—the optical pumping rate exceeds the decay rate of the lower level, where $\gamma$ is the transition frequency between the upper and lower sublevels, $\Gamma$ is the decay rate of the lower level, and S is the number of transitions induced by a radiation flux of 1 erg $cm^{-2}\ s^{-1}$ at the wavelength of the transition in question (e.g., 1083 nm for the metastable-helium line).

The last condition effectively means that the system must retain atoms within a cycle of excitation and relaxation via resonant radiation for several tens of consecutive events before the system is disrupted by a transition from either the upper or the lower level to other levels. Such leakage can occur both through direct photoexcitation of the aligning lower level to other, higher levels and through photoexcitation after a transition to the upper level of the cycle under consideration.

Because this paper considers UV wavelengths, which are often weakly represented in the radiation spectra of many stars, a detailed account of the conditions governing the population of the level from which the alignment-sensitive triplet line originates is particularly important. We have therefore expanded, relative to [9], the list of conditions underlying the selection of candidate lines for observations of exoplanetary magnetic fields with the Spektr-UV telescope:

- The decay rate of the upper levels back to the aligning level must exceed the population-loss rate through transitions to other lower-lying levels. In other words, the system must retain atoms within the excitation–relaxation cycle for several tens of consecutive events.

- No substantial population leakage to higher-lying levels. Such leakage can occur both through direct photoexcitation of the aligning level to other, higher levels and through photoexcitation after a transition to an upper level of the cycle under consideration. Therefore, the excitation rate into the upper levels considered must dominate over excitation to higher-lying states, and the upper levels must decay predominantly back to the aligning level.
- A sufficiently large population of the aligning level itself. In general, this population may be maintained by photoexcitation from lower-lying levels or by relaxation from higher-lying levels, but determining it requires a full nonequilibrium calculation of the entire set of atomic levels. Therefore, the method considered assumes that the aligning level already has a sufficient population, and the principal calculation is aimed at determining the conditions under which alignment is preserved.

Thus, the sensitivity of a particular line to atomic alignment is determined not only by the existence of conditions for alignment but also by whether the population is retained during excitation–relaxation cycles.

## 3. RESULTS

### 3.1 Selection of elements and lines

In accordance with the criteria described in the previous section, and based on the operating wavelength range and spectral resolution of the Spektr-UV telescope's spectrographs (115–310 nm at a resolution of 50,000 and 115–305 nm at a resolution of 1,000), we selected triplet lines of heavy elements in the UV range that are sensitive to alignment. Table 1 lists the parameters of the selected triplet transitions that are potentially suitable for diagnosing magnetic fields through atomic alignment. The complete list of lines is presented in Appendix 1. In column order, the table gives the ion name, the required number of alignment cycles Q_cyc, the effective cycling rate R_cyc (including the times for photoexcitation and radiative decay), the vacuum wavelengths λ of the triplet transitions, the energies of the lower and upper levels in eV, and the total angular momentum of the ion J_g. The final two columns give the ratios of sublevel statistical weights relative to the first transition in the triplet for the equilibrium and aligned cases, respectively. We note that some rows of the table contain Q_cyc → ∞ because the ground level is aligned and there are no radiative decays from it (although alignment can still be destroyed by electron collisions).

When calculating coefficients characterizing the efficiency of atomic alignment, only transitions induced by anisotropic stellar radiation and the corresponding radiative relaxation channels are included. Collisional processes are not included directly in these coefficients. Their influence is taken into account separately using local atmospheric parameters obtained from the three-dimensional hydrodynamic model described in Section 2.1; these parameters are used to determine the local collision frequencies of electrons with the atoms and ions under consideration. The preservation of alignment, taking account of the influence of collisions on level populations, is tested by comparing characteristic radiative cycling rates with the frequencies of collisional processes in the formation region of the relevant line.

The influence of the magnetic field is included at the stage of the quantum-mechanical calculation of atomic alignment, following the methodology described in detail in [9]. In this formulation, the magnetic field changes the distribution of magnetic-sublevel populations and thus the degree of alignment and the relative populations of multiplet levels, owing to rotation of the quantization axis and disruption of anisotropy in the coordinate system associated with the directed radiation flux. For weak magnetic fields, the influence of the field on the hydrodynamic structure of the atmosphere itself may be neglected: the calculations show that the characteristic distributions change very little for field strengths up to 1 G. Therefore, in the cases considered in this paper, the field is treated as a factor affecting the quantum state of atoms and ions, while the atmospheric structure remains unchanged.

The limited number of elements included in Table 1 is based on the results of previous observations described in the Introduction. Transit absorption by hot exoplanets is known to have been observed in Ca II, Fe I, Fe II, Mg I, Mn I, O I, Mg II, and Sc II lines; consequently, our predictions are limited to the same elements and to transitions with characteristics similar to those observed previously. For example, the energies of the lower levels of the transitions rarely exceed 10 eV; in compiling the table, we took this constraint into account, together with the magnitude of the radiation flux in the resonance region.

From the full list of identified transitions, the lines of greatest interest for future observations with Spektr-UV were selected for Table 1. Priority was given to transitions that simultaneously satisfy several conditions: high efficiency of the repeated excitation–relaxation cycle, characterized by high Q_cyc and R_cyc values; sufficiently low energy of the aligning level, increasing the likelihood of a substantial population in the upper atmosphere; a significant deviation of the relative statistical weights of the triplet components from the equilibrium 2J+1 distribution; and sufficient spectral separation of the components to allow their transit absorption to be measured independently. Additional preference was given to atoms and ions whose presence has already been established in the atmospheres of hot exoplanets, since detectable signals are considerably more likely for such transitions. Thus, the lines listed below should be regarded as the most promising compromise between the efficiency of atomic alignment and the practical feasibility of detecting its spectroscopic signature.

**Table 1. Highest-priority lines for Spektr-UV that are subject to the alignment effect.**

| Element | Q_cyc | R_cyc, $s^{-1}$ | λ, Å | Energy, eV | J_g | Statistical weights without alignment | Statistical weights with alignment |
|---|---|---|---|---|---|---|---|
| HAT-P-32 | | | | | | | |
| Fe II | 1055 | 233 | 2599.1<br>2612.7<br>2626.5 | 0.048−4.82<br>0.048−4.79<br>0.048−4.77 | 3.5 | 1:1.33:1.66 | 1: 1.22: 1.63 |
| Fe I | 2425 | 144 | 2525.1<br>2530.6<br>2541.7 | 0.110−5.02<br>0.110−5.01<br>0.110−4.98 | 1.0 | 1:3:5 | 1:2.17: 4.17 |
| Fe II | 6950 | 850 | 2710.2<br>2725.7<br>2750.1 | 1.040−5.62<br>1.040−5.59<br>1.040−5.55 | 2.5 | 1:1.5:2 | 1:1.35: 1.94 |
| KELT-20 | | | | | | | |
| Fe I | 3826 | 227 | 2525.1<br>2530.6<br>2541.7 | 0.110−5.02<br>0.110−5.01<br>0.110−4.99 | 1.0 | 1:3:5 | 1:2.17: 4.17 |
| Fe II | 4567 | 721 | 2328.1<br>2349.0<br>2381.5 | 0.083−5.41<br>0.083−5.36<br>0.083−5.29 | 2.5 | 1:1.5:2 | 1:1.35: 1.94 |
| Fe II | 1055 | 227 | 2599.1<br>2612.7<br>2626.5 | 0.048−4.82<br>0.048−4.79<br>0.048−4.77 | 3.5 | 1:1.33:1.66 | 1: 1.22: 1.63 |

We introduced several key quantitative parameters describing the efficiency with which population anisotropy is induced for each triplet transition. Let us designate the energy levels (Fig. 1) as follows: a is the aligning level; u_i are the upper levels through which the excitation–relaxation cycles proceed; c is any level above u_i; and d is any level below a.

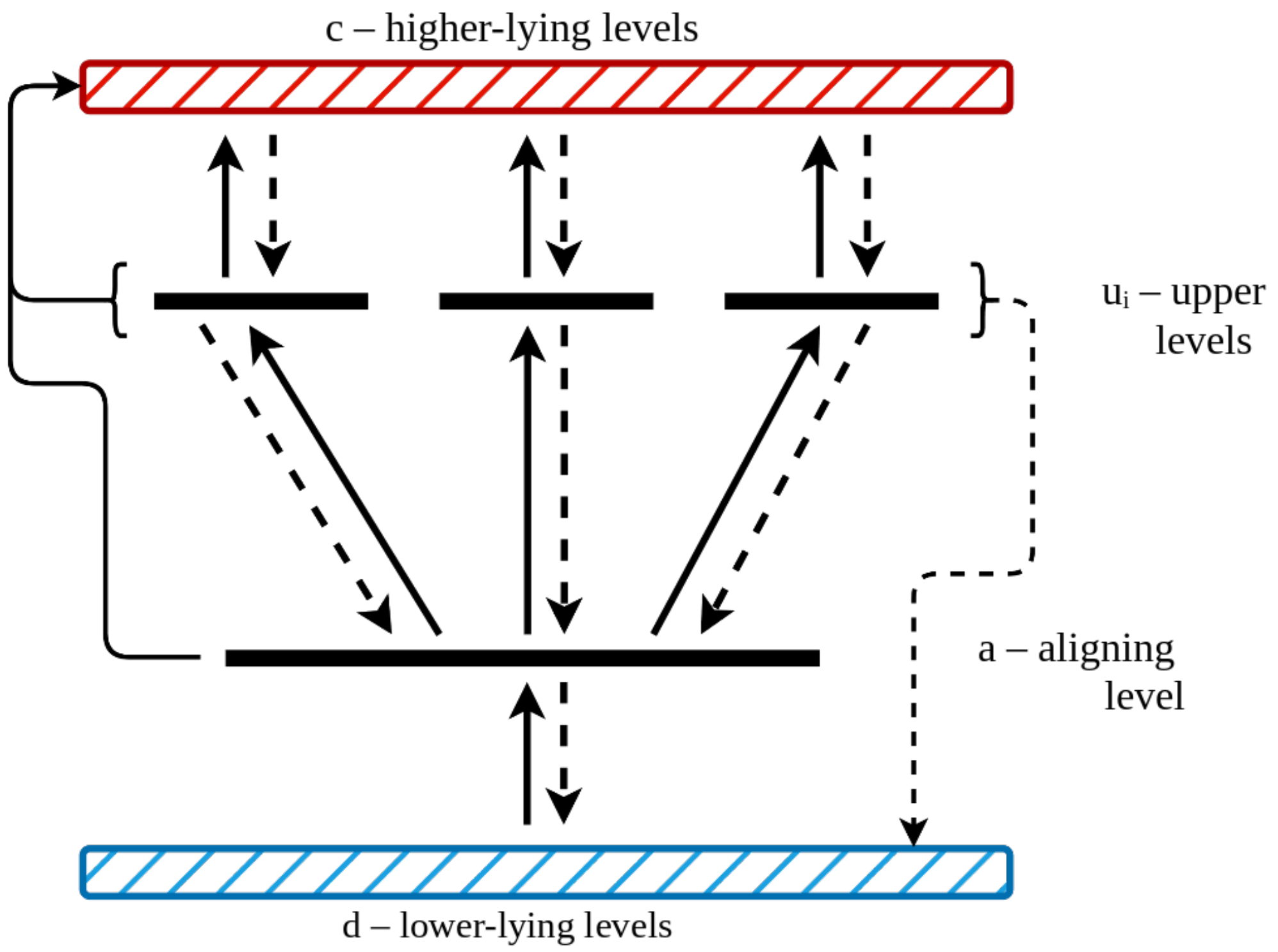

Fig. 1. Level scheme for the alignment effect. Solid arrows indicate excitation pathways, and dashed arrows indicate relaxation pathways. Higher- and lower-lying levels refer to the entire set of levels of the given ion with energies above or below those of the system under consideration.

To test the criteria, the photoexcitation rate R was calculated from stellar spectra scaled to the planet's orbital distance. We define the total photoexcitation rate R_exc as the sum over the three transitions from a to u_i, and the decay rate Γ_a of the aligning level to all lower-lying levels d. The branching ratios β_i are also useful quantities, indicating the fraction of atoms that return to the initial aligning level a after excitation. A small $\beta_i$ corresponds to population leakage into other lower levels d and, consequently, to weakening of the quasi-isolated cycle.

$$\beta_i = \frac{A(u_i \to a)}{\sum_k A(u_i \to k)} \tag{1}$$

where A denotes the Einstein coefficients.

The efficiency of repeated passage through the a → $u_i$ → a cycle is most directly characterized by $R_{cyc}$—the effective cycling rate, which takes account of the probabilities $\beta_i$ of returning to the initial level:

$$R_{cyc} = \sum_i R(a \to u_i)\beta_i \tag{2}$$

Consequently, the number of alignment cycles $Q_{cyc}$ undergone by an ion during the lifetime of the aligning level is $Q_{cyc} = R_{cyc}/\Gamma_a$. Thus, large $Q_{cyc}$ and $R_{cyc}$ values indicate that the populations of magnetic sublevels can be redistributed repeatedly before depolarization or loss from the system, making the corresponding transition physically more promising for observing magnetic sensitivity. Additional selection criteria are the suppression of upward population leakage to higher levels c and the requirement that radiative return $u_i$ → a, rather than further excitation, dominate after excitation a → $u_i$. The quantities presented should therefore be regarded not only as atomic parameters of individual lines but also as characteristics of a coupled level-kinetics system determining how efficiently radiative pumping can maintain population anisotropy and thereby produce an observable atomic-alignment effect.

### 3.2 Modeling the selected lines with the three-dimensional model

To predict the observability of the selected absorption lines, we used calculations whose parameters had previously been chosen by comparing three-dimensional aeronomic simulations with the observed transit spectra of KELT-20 b and HAT-P-32 b [43,44]. Thus, the systems considered represent two substantially different upper-atmosphere regimes: the compact, nearly hydrostatic atmosphere of KELT-20 b and the extended, escaping atmosphere of HAT-P-32 b.---

For KELT-20 b, we used the solution that provided the best simultaneous reproduction of the observed Balmer-series and Fe II 5018 Å lines. In the original calculations, the planet was modeled using the upper observational estimate of its mass, $M_p = 3.52\ M_{Jup}$, and a radius $R_p = 1.83\ R_{Jup}$. KELT-20 b orbits an A2V star of mass 2 $M_{Sun}$, radius 1.561 $R_{Sun}$, and temperature 8981 K at a distance of 0.0542 au. At an assumed lower-boundary equilibrium temperature of 2200 K, the planet's strong gravity leads to a very compact atmosphere: the characteristic scale height is approximately 0.005 $R_p$, and the main contribution to H I and Fe II absorption comes from below an altitude of about 0.3 $R_p$ above the photometric radius at ~0.1 bar. The computational domain was limited to $R_{max} = 4\ R_p$ with a spatial step of approximately 0.002 $R_p$. The influence of the stellar wind in the formation region of the lines considered was neglected.

In this work, the baseline solution chosen for KELT-20 b combines a metallicity [Fe/H] = −1 with a vacuum-ultraviolet (VUV, 91.2–300 nm) flux $F_{VUV} = 3\times10^7$ erg cm$^{-2}$ s$^{-1}$ (scaled to a distance of 1 au); this combination provided a consistent reproduction of four Balmer-series lines and Fe II 5018 Å. In this calculation, the fluxes $F_{XUV}$ (extreme ultraviolet, 0.1–91.2 nm) and $F_{Ly\alpha}$ (121.6 nm) were both set to 10 erg cm$^{-2}$ s$^{-1}$ at 1 au. These values were chosen arbitrarily because an analysis of the influence of these fluxes on atmospheric structure and observed absorption showed that they have no effect on it. This is because the large flux in the VUV region suppresses all observable effects of the other two fluxes (heating and ionization degree). The pressure at the lower boundary was set to 0.1 bar. Owing to the high temperature, the atmosphere was assumed to be predominantly atomic and included hydrogen and the heavy elements C, O, Mg, Fe, and Na. This solution is characterized by a hydrostatic regime with a very low mass-loss rate, of order $10^4$ g s$^{-1}$.

HAT-P-32 b is a hot Jupiter of mass 0.75 $M_J$ and radius 1.789 $R_J$, orbiting at a distance of 0.03427 au around an F9-type star of mass 1.176 $M_{Sun}$, radius 1.387 $R_{Sun}$, and temperature 6000 K, which yields a planetary equilibrium temperature of about 2000 K. For this planet, a fundamentally different type of solution was obtained. Here, the atmosphere is highly extended and undergoes intense hydrodynamic outflow. When modeling profiles near mid-transit, the computational domain was limited to $R_{max} = 10\ R_p$; the stellar wind was not included in this local formulation. We note that the effective radius of the Roche lobe is approximately 1.5 $R_p$, and the distance to the L1 Lagrange point is 2.1 $R_p$. A molecular atmosphere containing heavy elements and having solar metallicity was used as the baseline. Metals have a substantial effect on the structure of HAT-P-32 b's atmosphere: their photoionization increases the electron density in the dense layers, accelerates $H_2$ dissociation, and leads to the formation of a more extended envelope.

The specific parameter set for HAT-P-32 b corresponds to model No. 24 in [44] for a molecular atmosphere with metals: [Fe/H] = 0, an H/He ratio of 98/2, $F_{HXR} = 80$ erg cm$^{-2}$ s$^{-1}$ (hard X-rays, 0.5–10 nm), $F_{Ly\alpha} \approx 290$ erg cm$^{-2}$ s$^{-1}$ at 1 au, and a lower-boundary temperature $T_{base} = 2000$ K. In the series of calculations, the VUV radiation flux was set to $7\times10^4$ erg cm$^{-2}$ s$^{-1}$ at 1 au. The lower boundary was located at $R_{base} = 1.1\ R_p$ with a pressure $P_{base} = 0.1$ bar. The calculated mass-loss rate for this solution is $1.36\times10^{13}$ g s$^{-1}$.

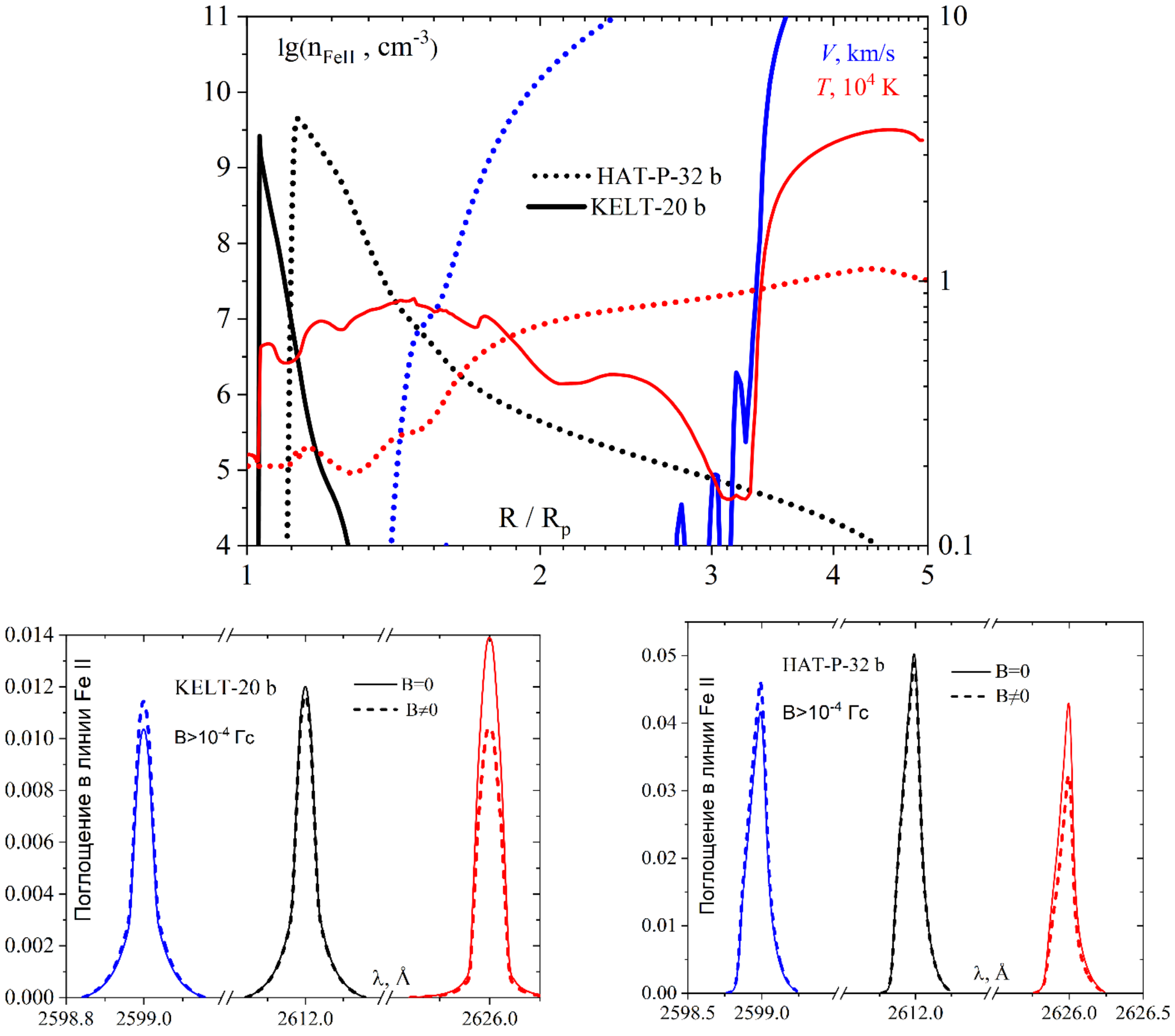

Fig. 2. Upper panel: profiles of Fe II ion density, plasma temperature, and velocity along the planet–star direction for the two planets considered.

Lower panel: calculated absorption profiles of the Fe II triplet components at 2599.1, 2612.7, and 2626.5 Å for KELT-20 b (left) and HAT-P-32 b (right). Dashed and solid lines show calculations with and without a magnetic field, respectively.

On the basis of the density, temperature, and velocity distributions (e.g., the upper panel of Fig. 2), the Fe II level populations were then calculated, and absorption was computed along optical rays directed from the stellar disk toward the planet. For each hydrodynamic solution, two variants of the magnetic-sublevel populations were constructed: with atomic alignment preserved (Fig. 2, lower panel, thin line) and with alignment destroyed by a magnetic field (Fig. 2, lower panel, thick line). The gas-dynamic structure of the atmosphere remained unchanged: the difference between the two types of spectra results from the redistribution of magnetic-sublevel populations and the corresponding change in the relative absorption coefficients of the triplet components.

It should be noted that the calculated deviation from the equilibrium distribution in the absence of a magnetic field is sufficiently large to manifest itself not only in the populations of atomic states but also directly in the transit spectrum. Atomic alignment changes the relative depths of the triplet components by amounts of the order of tenths of a percentage point or more. These numbers should not be interpreted directly as an exact measurement of magnetic-field strength. The method is most sensitive in the transition region of atomic alignment; at fields of the order of several to ten gauss, the effect reaches saturation. Thus, analysis of absorption spectra provides, rather, a lower field threshold of $B = 2\times10^{-5}$ G. The presence of alignment indicates that the field does not exceed this value. The absence of alignment, if caused by the presence of a field, indicates that the field must be greater than this value.

For KELT-20 b, the most pronounced change is observed in the 2626 Å component: its depth decreases from 1.39% in the absence of a magnetic field to 1.04% in the calculation with a field. The 2599 Å component changes in the opposite direction—its depth increases from 1.04% to 1.14%—whereas the 2612 Å component changes only slightly, from 1.20% to 1.17%. Thus, the presence of a magnetic field does not uniformly weaken or strengthen the entire triplet but causes a characteristic redistribution of the relative depths of its components. A similar rearrangement of the component ratios is observed for HAT-P-32 b, although the absolute line depths are considerably greater because of the more extended absorption region (Fig. 3, left). The depth of the 2626 Å component changes from 4.30% to 3.23%, that of the 2599 Å component from 4.17% to 4.60%, and that of the 2612 Å component from 5.03% to 4.89%. In both cases, the differences between the spectra obtained with and without alignment are sufficiently large to be distinguished against the noise, provided that the observing parameters are chosen appropriately (Section 3.3).

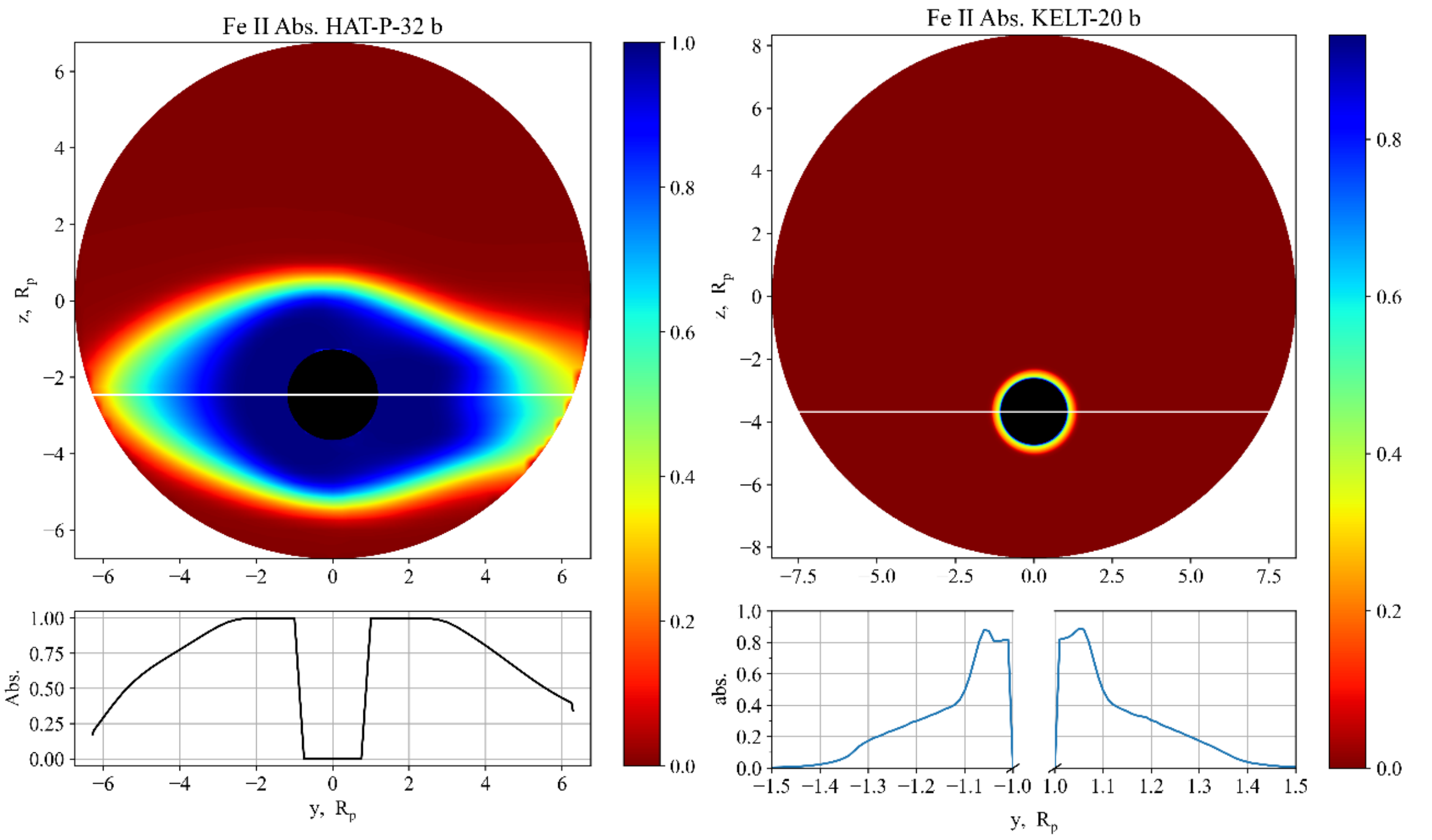

Fig. 3. Spatial absorption maps for HAT-P-32 b and KELT-20 b near the central triplet line in the lower panel of Fig. 2, λ = 2612 Å. The maps show which parts of the escaping atmosphere make the principal contribution to the transit signal.

For the selected Fe II triplet, the effective optical cycling rates in the HAT-P-32 and KELT-20 systems are $R_{cyc} = 233$ and $227\ s^{-1}$, respectively. These were compared with distributions of local collision frequencies obtained using the calculated gas-density and temperature profiles in each model cell. For Fe II interactions with neutral hydrogen, we used quantum-mechanical data on inelastic processes from [45], $\langle\sigma v\rangle \approx 2.3\times10^{-9}\ cm^3\ s^{-1}$ and $\sigma \approx 4.5–5\times10^{-15}\ cm^2$ at $T_H \sim 0.1$ eV. For electron collisions, we used effective cross sections from [46], yielding coefficients of order $10^{-8}–10^{-7}\ cm^3\ s^{-1}$ for low-lying transitions at $T_e \approx 1000$ K.

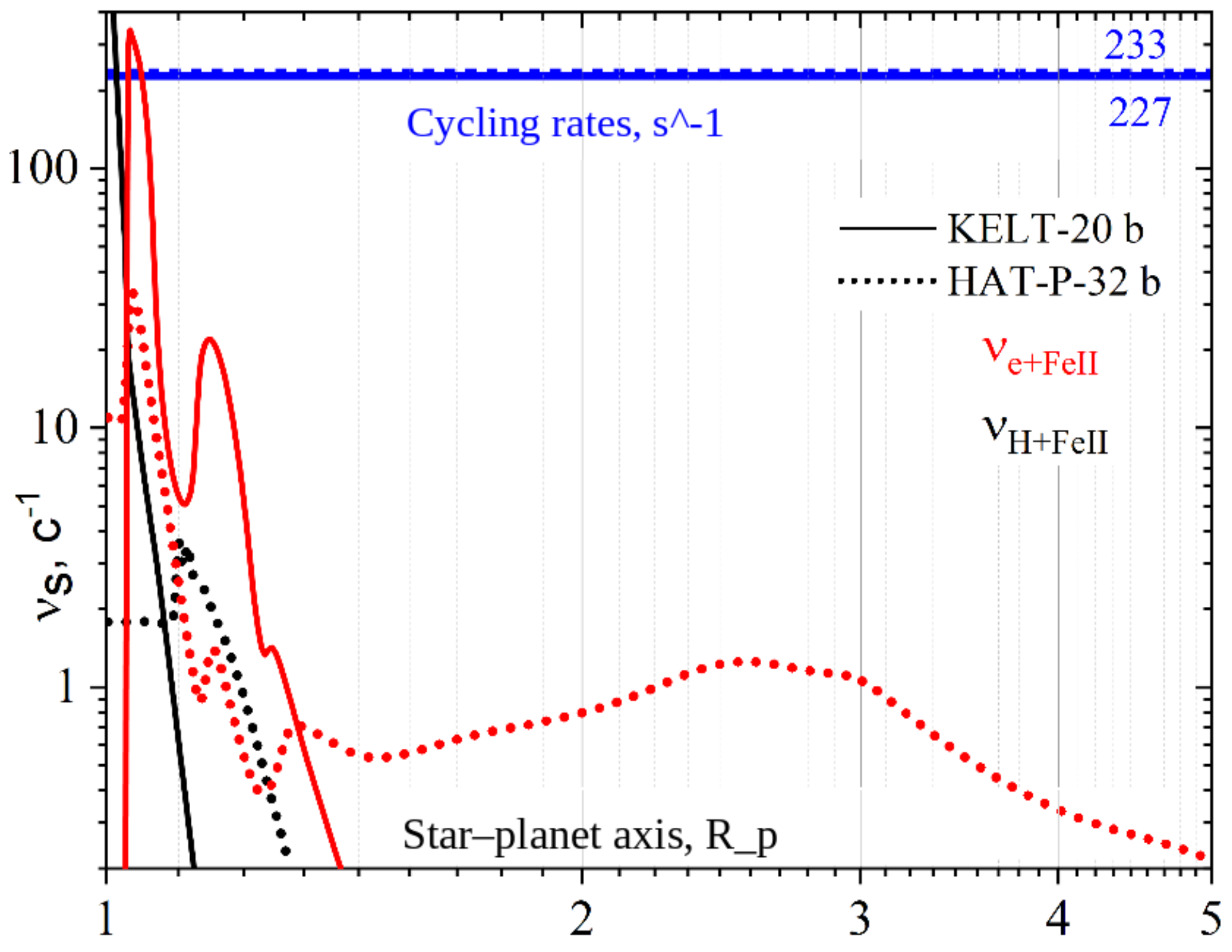

Fig. 4. Collision frequencies of Fe II with hydrogen atoms (black solid and dashed lines) and with electrons (red solid and dashed lines) along the star–planet axis. Dashed lines indicate the triplet cycling rates, R_cyc ≈ 227 $s^{-1}$ for KELT-20 b and 233 $s^{-1}$ for HAT-P-32 b.

Comparison of the collision rates and optical cycling frequency does not allow the entire absorption region to be regarded as collisionless, as is required to preserve alignment in the selected triplet. Although the collision frequency of iron ions with electrons is insignificant, hydrogen atoms, which dominate planetary atmospheres, may substantially affect Fe II level kinetics. Near the lower atmospheric boundary, the Fe II–H collision frequency is comparable to or exceeds R_cyc, but, owing to the rapid decrease in density, becomes much smaller than R_cyc beyond approximately 1.1–1.2 R_p. Figure 3 shows that absorption in the triplet lines occurs on spatial scales much larger than the collisional region; therefore, collisions can be regarded as having no substantial influence in the Fe II triplet absorption-formation region and consequently cannot destroy the alignment effect.

### 3.3. Exposure-time calculations for Spektr-UV

We assessed the possibility of observing the alignment effect with the Spektr-UV telescope using the telescope's online exposure-time calculator [39]. We considered the high-resolution

UV spectrograph operating over 174–310 nm with R ≈ 50,000. One resolution element then corresponds to $\Delta v \approx c/R \approx 6$ km s$^{-1}$, or about 2 km s$^{-1}$ per pixel with three pixels per resolution element.

We first obtain an estimate that does not depend on accounting for absorption in the interstellar medium. For a line depth of 1.5%—the value predicted by Exo3D model calculations for KELT-20 b—the signal significance in a single resolution element is approximately 0.015 SNR_transmission (SNR_transmission, or SNR_tr, is the signal-to-noise ratio of the final transmission spectrum after combining in-transit and out-of-transit spectra). Thus, detection at the 3σ and 5σ levels requires (S/N)_tr ≈ 200 and 333, respectively. For the ratio of the averaged in-transit and out-of-transit spectra, we used the expression

$$\left( S/N \right) = s\sqrt{\frac{N_{in} N_{out}}{N_{in} + N_{out}}} \tag{3}$$

where s is the signal-to-noise ratio of a single exposure per resolution element, and N_in and N_out are the numbers of in-transit and out-of-transit exposures. For a fixed total observing time, the highest S/N is obtained when N_in and N_out are comparable; therefore, a long series of out-of-transit exposures alone does not compensate for a shortage of in-transit frames.

The duration of an individual exposure is limited by changes in the planet's radial velocity. The spectrograph accumulates photons throughout the exposure, during which the planet moves along its orbit, so the absorption lines of the planetary atmosphere undergo Doppler shifts. Smeared lines become broader and shallower. If the planet's velocity at the start of an exposure is taken as V_p(t = t_start) and at the end as V_p(t = t_end), then Δv_sm = |V_p(t_start) − V_p(t_end)|. If the velocity changes approximately linearly, then Δv_sm ≈ (dV_p/dt)×t_exp (from a Taylor-series expansion about $t_0$, the midpoint of the exposure). For a circular orbit, V_p(t) = K_p sin[2π(t − $T_0$)/P], or dV_p/dt = (2πK_p/P)×cos[2π(t − $T_0$)/P] (where K_p is the radial-velocity semi-amplitude of the planet, P is its orbital period, and $T_0$ is the mid-transit time). At mid-transit, t ≈ $T_0$ and cos(0) = 1, giving dV_p/dt = 2πK_p/P. Substituting this into Δv_sm gives Δv_sm ≈ (2πK_p/P)×t_exp (where t_exp is the duration of one spectral exposure). Thus, near mid-transit, smearing can be estimated as Δv_sm ≈ (2πK_p/P)×t_exp.

For KELT-20 b the smearing coefficient is $3.62\times10^{-3}$ km s$^{-2}$, and for HAT-P-32 b it is $5.81\times10^{-3}$ km s$^{-2}$. The constraint Δv_sm = 2–3 km s$^{-1}$ corresponds to 340–520 s and 550–830 s, respectively. We therefore recommend a 300 s exposure for KELT-20 b, whereas 400–500 s is more practical for HAT-P-32 b. A 600 s exposure for HAT-P-32 b already smears the profile by approximately 3.5 km $s^{-1}$ and barely increases the final S/N at unchanged total observing time.

To calculate the required exposure times, we used stellar radiation fluxes taken for KELT-20 from [47] and for HAT-P-32 from the authors' archival data. In a test calculation, a single 300 s exposure of KELT-20 yields s ≈ 201. During a transit lasting 3.57 h, approximately 43 exposures can be obtained both in and out of transit, yielding $(S/N)_{tr} \approx 931$. For HAT-P-32 b, a 600 s exposure gives s ≈ 15 and 18 exposures during each 3.12 h interval, i.e., $(S/N)_{tr} \approx 45$. Switching to 400–500 s preserves almost the same final value but reduces velocity smearing and provides more points on the transit light curve.

**Table 2. Attenuation of UV flux by interstellar material for HAT-P-32 b and KELT-20 b. Each cell gives calculations for HAT-P-32 b / KELT-20 b.-**

| λ, nm | A(λ), mag | T = 10−0.4A(λ) | Flux loss, % |
|---|---|---|---|
| 174 / 174 | 0.414 / 0.157 | 0.68 / 0.86 | 32 / 24 |
| 200 / 200 | 0.470 / 0.170 | 0.649 / 0.85 | 35 / 25 |
| 215.5 / 214.3 | 0.527 / 0.200 | 0.615 / 0.83 | 39 / 27 |
| 250 / 250 | 0.3821 / 0.145 | 0.703 / 0.87 | 30 / 23 |
| 300 / 300 | 0.300 / 0.140 | 0.758 / 0.90 | 25 / 10 |

Absorption by the interstellar medium was also taken into account separately using the three-dimensional maps from [48]. We obtained $A_{550} = 0.0629$ mag for KELT-20 and 0.1652 mag for HAT-P-32, where $A_{550}$ is the interstellar extinction at a wavelength of 550 nm, expressed in magnitudes, and indicates how much interstellar dust attenuates the flux at that wavelength. To calculate extinction at another wavelength, one needs the interstellar extinction curve, which is determined empirically from stellar observations. For example, extinction at UV wavelengths can be calculated using the CCM89 law [49], which gives

$$\frac{A_\lambda}{A_V} = a(x) + \frac{b(x)}{R_V}, \tag{4}$$

where $x = 1/\lambda[\mu m]$ and $R_V = A_V/E(B-V) = 3.1$ is the total-to-selective extinction ratio; its standard value for the diffuse interstellar medium is 3.1. Thus, to calculate extinction at any wavelength, the following equation should be solved:

$$A_\lambda = A_V\left(a(x)+\frac{b(x)}{R_V}\right). \tag{5}$$

The wavelength range λ = 174–310 nm corresponds to x = 3.23–5.75 μm$^{-1}$. According to CCM89, the following UV formulae should be used for 3.3 ≤ x ≤ 8.0:

$$a(x) = 1.752 - 0.316x - \frac{0.104}{(x-4.67)^2+0.341} + F_a(x) \tag{6}$$

$$b(x) = -3.090 + 1.825x + \frac{1.206}{(x-4.62)^2+0.263} + F_b(x). \tag{7}$$

For x < 5.9, F_a(x) = F_b(x) = 0. Thus, according to the CCM89 law [41], the transmission in the region of the 2599–2627 Å triplet is 0.883–0.885 and 0.721–0.724, respectively. Because photon S/N scales as the square root of transmission, estimates of the S/N of the final absorption spectrum per transit per resolution element, corrected for interstellar absorption, decrease to ≈875 and ≈38 for KELT-20 b and HAT-P-32 b, respectively. The interstellar-absorption correction should be applied only if the exposure-time calculator was supplied with the original model stellar flux, unattenuated by the interstellar medium. If instead the calculator was supplied with the observed catalog flux, which already includes interstellar absorption, no additional correction should be applied, as this would double-count the extinction.

The calculations show that detecting the predicted atomic-alignment effect in KELT-20 b with Spektr-UV appears entirely realistic within a single transit, whereas observing HAT-P-32 b is considerably more difficult because of its lower signal-to-noise ratio. When planning the observations, it is crucial to choose exposure times that account for Doppler smearing of the lines and to account correctly for interstellar extinction. The resulting estimates make it possible to formulate optimal parameters for an observing program aimed at testing the model predictions.

## 4. CONCLUSION

This paper proposes a systematic procedure for selecting UV spectral lines suitable for searching for exoplanetary magnetic fields using the atomic-alignment method, which is based on deviations from the equilibrium ratio of absorption-line depths within certain triplets. Predictions of the detectability of absorption lines are based on the numerical global three-dimensional multi-fluid hydrodynamic model Exo3D [39–41], which describes effects important for atmospheric formation. The redistribution of magnetic fine-structure sublevel populations by an anisotropic radiation spectrum in the presence or absence of magnetic fields at each local point in the atmosphere is calculated using the method described in detail in [9]. In this work, the calculation of the alignment effect was supplemented with a more detailed treatment of an exhaustive set of conditions for preserving alignment of the lower levels of individual triplets. The most suitable lines selected are given in Table 1 of Section 3.1 and in the Appendix (87 lines). The calculations show that the transition from preserved atomic alignment to alignment destroyed by a magnetic field is accompanied by a change in the ratio of line amplitudes within the selected triplet. One iron-ion triplet is examined in detail; the difference between its two calculated spectra is approximately one third of the full line amplitude for both exoplanets considered, KELT-20 b and HAT-P-32 b. The changes in different lines within the triplet are not equal in magnitude, so joint analysis of all three lines appears more informative than measuring the depth of a single component. Comparing collision frequencies with the optical cycling rate allows us to estimate where in the atmosphere atomic alignment can be preserved. In the extended atmosphere of HAT-P-32 b, a substantial fraction of the absorption is formed outside the dense lower layers, where the collision frequency exceeds the cycling rate, so collisions have little influence on the calculated component ratios there. In the more compact atmosphere of KELT-20 b, the $Fe^{+}$ formation region overlaps more strongly with the inner layers, where collisions with hydrogen atoms may partly modify the magnetic-sublevel populations. For this reason, the contrast obtained for KELT-20 b should be regarded as a model estimate to be refined in subsequent work.

The exposure-time calculations, consistent with the modeled profiles, make KELT-20 b the primary target for the first observing program. With 300 s exposures, one ideal transit yields a significance of approximately 2.8–3.0σ; therefore, a realistic minimum is two transits, and approximately four are needed to reach 5σ. For HAT-P-32 b, the line amplitudes are larger, but the star is much fainter in the UV: statistically, 21–28 transits are required for 3σ and 56–77 for 5σ. It is more reasonable to regard this object as a second stage of the program or as a target for jointly accumulating data over several observing cycles.

**FUNDING**

This work was supported by Russian Science Foundation grant No. 25-72-10080.

## REFERENCES

1. *P. Zarka*, Planet. Space Sci. **55**, 598 (2007).

2. *C. R. Lynch, T. Murphy, E. Lenc, and. Kaplan*, Mon. Not. R. Astron. Soc. **478**, 1763 (2018).

3. *J.-M. Grießmeier, P. Zarka, and H. Spreeuw*, Astron. Astrophys. **475**, 359 (2007).

4. *M. P. van Haarlem, M. W. Wise, A. W. Gunst, G. Heald, et al.*, Astron. Astrophys. **556**, A2 (2013).

5. *J. D. Turner, P. Zarka, J.-M. Grießmeier, J. Lazio, et al.*, Astron. Astrophys. **645**, A59 (2021).

6. *J. D. Turner, J.-M. Grießmeier, P. Zarka, X. Zhang, and E. Mauduit*, Astron. Astrophys. **688**, A123 (2024).

7. *P. W. Cauley, E. L. Shkolnik, J. Llama, and A. F. Lanza*, Nat. Astron. **3**, 1128 (2019).

8. *A. Oklopčić, M. Silva, P. Montero-Camacho, and C. M. Hirata*, Astrophys. J. **890**, 62 (2020).

9. *M.S. Rumenskikh, A. V. Taichenachev, I. F. Shaikhislamov, V. I. Yudin, and E. D. Savintseva*, Mon. Not. R. Astron. Soc. **541**, 2872 (2025).

10. *D. A. Varshalovich*, Sov. Phys. Usp. **13**, 1 (1970).

11. *T. Azevedo Silva, O. D. S. Demangeon, N. C. Santos, R. Allart, et al.*, Astron. Astrophys. **666**, L10 (2022).

12. *M. Ben-Yami, N. Madhusudhan, S. H. C. Cabot, S. Constantinou, et al.*, Astrophys. J. Lett. **897**, L5 (2020).

13. *H. J. Hoeijmakers, D. Ehrenreich, K. Heng, D. Kitzmann, et al.*, Nature **560**, 453 (2018).

14. *H. J. Hoeijmakers, D. Ehrenreich, D. Kitzmann, R. Allart, et al.*, Astron. Astrophys. **627**, A165 (2019).

15. *V. Bourrier, D. Ehrenreich, M. Lendl, M. Cretignier, et al.*, Astron. Astrophys. **635**, A205 (2020).

16. *S. H. C. Cabot, N. Madhusudhan, L. Welbanks, A. Piette, and S. Gandhi*, Mon. Not. R. Astron. Soc. **494**, 363 (2020).

17. *M. Stangret, L. Fossati, M. C. D'Arpa, F. Borsa, et al.*, Astron. Astrophys. **692**, A76 (2024).

18. *A. B. Langeveld, N. Madhusudhan, and S. H. C. Cabot*, Mon. Not. R. Astron. Soc. **514**, 5192 (2022).

19. *J. D. Turner, E. J. W. de Mooij, R. Jayawardhana, M. E. Young, et al.*, Astrophys. J. Lett. **888**, L13 (2020).

20. *F. Yan, N. Casasayas-Barris, K. Molaverdikhani, F. J. Alonso-Floriano, et al.*, Astron. Astrophys. **632**, A69 (2019).

21. *H. M. Tabernero, M. R. Zapatero Osorio, R. Allart, F. Borsa, et al.*, Astron. Astrophys. **646**, A158 (2021).

22. *N. Casasayas-Barris, E. Palle, M. Stangret, V. Bourrier, et al.*, Astron. Astrophys. **647**, A26 (2021).

23. *L. Fossati, F. Biassoni, G. M. Cappello, F. Borsa, et al.*, Astron. Astrophys. **676**, A99 (2023).

24. *F. Borsa, L. Fossati, T. Koskinen, M. E. Young, and D. Shulyak*, Nat. Astron. **6**, 226 (2022).

25. *S. Czesla, M. Lampón, J. Sanz-Forcada, A. García Muñoz, et al.*, Astron. Astrophys. **657**, A6 (2022).

26. *S. R. Merritt, N. P. Gibson, S. K. Nugroho, E. J. W. de Mooij, et al.*, Astron. Astrophys. **636**, A117 (2020).

27. *S. R. Merritt, N. P. Gibson, S. K. Nugroho, E. J. W. de Mooij, et al.*, Mon. Not. R. Astron. Soc. **506**, 3853 (2021).

28. *D. K. Sing, P. Lavvas, G. E. Ballester, A. Lecavelier des Etangs, et al.*, Astron. J. **158**, 91 (2019).

29. *M. S. Rumenskikh, M. L. Khodachenko, I. F. Shaikhislamov, I. B. Miroshnichenko, A. G. Berezutsky, A. V. Shepelin, and N. K. Dwivedi*, Mon. Not. R. Astron. Soc. **526**, 4120 (2023).

30. *J. L. Linsky, H. Yang, K. France, C. M. Johns-Krull, A. Brown, and G. J. Herczeg*, Astrophys. J. **717**, 1291 (2010).

31. *A. Vidal-Madjar, A. Lecavelier des Etangs, J.-M. Desert, G. E. Ballester, R. Ferlet, G. Hebrard, and M. Mayor*, Astrophys. J. **676**, L57 (2008).

32. L. *Ben-Jaffel*, Astrophys. J. **671**, L61 (2007).

33. *L. Ben-Jaffel*, Astrophys. J. **688**, 1352 (2008).

34. *M. Holmstrom, A. Ekenback, F. Selsis, T. Penz, H. Lammer, and P. Wurz*, Nature **451**, 970 (2008).

35. *R. O. P. Loyd, K. France, A. Youngblood, C. Schneider, et al.*, Astrophys. J. **824**, 102 (2016).

36. *G. N. Tsurikov, D. V. Bisikalo, V. I. Shematovich, and A. G. Zhilkin*, Astron. Rep. **68**, 1406 (2024).

37. *I. F. Shaikhislamov, I. B. Miroshnichenko, S. S. Sharipov, M. S. Rumenskikh, et al.*, Astron. Astrophys. **696**, A211 (2025).

38. *A. G. Berezutsky, I. F. Shaikhislamov, M. S. Rumenskikh, I. B. Miroshnichenko, M. L. Khodachenko, M. P. Golubovskii, and S. S. Sharipov*, Mon. Not. R. Astron. Soc. **533**, 374 (2024).

39. *I. F. Shaikhislamov, M. L. Khodachenko, H. Lammer, A. G. Berezutsky, I. B. Miroshnichenko, and M. S. Rumenskikh*, Mon. Not. R. Astron. Soc. **481**, 5315 (2018).

40. *M. S. Rumenskikh, I. F. Shaikhislamov, M. L. Khodachenko, H. Lammer, I. B. Miroshnichenko, A. G. Berezutsky, and L. Fossati*, Astrophys. J. **927**, 238 (2022).

41. *M. L. Khodachenko, I. F. Shaikhislamov, H. Lammer, A. G. Berezutsky, et al.,* Astrophys. J. **885**, 67 (2019).

42. *A. Kramida, Yu. Ralchenko, and J. Reader*, NIST ASD Team, NIST Atomic Spectra Database (ver. 5.10) (National Institute of Standards and Technology, Gaithersburg, MD, 2022).

43. S. Sharipov, M. Rumenskikh, I. Shaikhislamov, I.B. Miroshnichenko, A.V. Shepelin, M. P. Golubovskii, and A.G. Berezutsky, Mon. Not. R. Astron. Soc. (in press).

44. *I. F. Shaikhislamov, S. S. Sharipov, M. L. Khodachenko, I. B. Miroshnichenko, M. S. Rumenskikh, M. P. Golubovsky, and A. G. Berezutsky*, arXiv:2603.14059 [astro-ph.CO] (2026).

45. *S. A. Yakovleva, A. K. Belyaev, and W. P. Kraemer*, Mon. Not. R. Astron. Soc. **483**, 5105 (2019).

46. *Y. Wan, C. Favreau, S. D. Loch, B. M. McLaughlin, Y. Qi, and P. C. Stancil*, Mon. Not. R. Astron. Soc. **485**, 2252 (2019).

47. *M. B. Lund et al*. AJ **154**, 194 (2017)

48. *J. L. Vergely, R. Lallement, and N. L. J. Cox*, Astron. Astrophys. **664**, A174 (2022).

49. *J. A. Cardelli, G. C. Clayton, and J. S. Mathis*, Astrophys. J. **345**, 245 (1989).

## APPENDIX. Table of UV lines sensitive to alignment

| Element | *Q_cyc* | *R_cyc*, $s^{-1}$ | λ, Å | Energy, eV | J_g | Statistical weights without alignment | Statistical weights with alignment |
|---|---|---|---|---|---|---|---|
| | | | | HAT-P-32 | | | |
| Fe II | 4226 | 304.7 | 2614.6<br>2621.2 | 0.107−4.85<br>0.107−4.84 | 1.5 | 1:2:3 | 1: 1.727: 2.8187 |
| | | | 2631.8 | 0.107−4.82 | | | |
| Fe II | 1502 | 237.4 | 2607.9<br>2618.4<br>2532.1 | 0.083−4.84<br>0.083−4.82<br>0.083−4.79 | 2.5 | 1:1.5:2 | 1:1.3523: 1.94097 |
| Fe II | 1081 | 232.6 | 2599.1<br>2612.7<br>2626.5 | 0.045−4.82<br>0.045−4.79<br>0.045−4.77 | 3.5 | 1:1.33:1.66 | 1: 1.21601: 1.6331 |
| Fe II | 6950 | 849.9 | 2710.2<br>2725.7<br>2750.1 | 1.040−5.62<br>1.040−5.59<br>1.040−5.55 | 2.5 | 1:1.5:2 | 1:1.3523: 1.94097 |
| Fe II | 6703 | 105.9 | 2396.1<br>2400.1<br>2405.6 | 0.083−5.26<br>0.083−5.25<br>0.083−5.24 | 2.5 | 1:1.5:2 | 1:1.3523: 1.94097 |
| Fe II | 6495 | 912.2 | 2693.6<br>2717.5<br>2756.6 | 0.986−5.59<br>0.986−5.55<br>0.986−5.48 | 3.5 | 1:1.33:1.66 | 1: 1.21601: 1.6331 |
| Fe II | 6349 | 136.6 | 2383.8<br>2389.4<br>2396.4 | 0.048−5.25<br>0.048−5.24<br>0.048−5.22 | 3.5 | 1:1.33:1.66 | 1: 1.21601: 1.6331 |
| Fe II | 3564 | 435.8 | 2728.3<br>2747.8<br>2773.5 | 1.040−5.58<br>1.040−5.55<br>1.040−5.51 | 2.5 | 1:1.5:2 | 1:1.3523: 1.94097 |
| Fe II | 2221 | 279.9 | 2737.8<br>2750.1<br>2769.8 | 1.076−5.60<br>1.076−5.58<br>1.076−5.55 | 1.5 | 1:2:3 | 1: 1.727: 2.8187 |
| Fe II | 1265 | 159.4 | 2567.7<br>2583.4<br>2611.9 | 1.076−5.90<br>1.076−5.88<br>1.076−5.82 | 1.5 | 1:2:3 | 1: 1.727: 2.8187 |
| Fe I | ∞ | 138.9 | 2913.0<br>2937.8<br>2967.8 | 0.000−4.26<br>0.000−4.22<br>0.000−4.18 | 4.0 | 1:1.2857: 1.5714 | 1: 1.17608: 1.544020 |
| Fe I | ∞ | 598.7 | 2448.5<br>2463.4<br>2484.0 | 0.000−5.06<br>0.000−5.03<br>0.000−4.99 | 4.0 | 1:1.2857: 1.5714 | 1: 1.17608: 1.544020 |
| Fe I | 2425 | 143.5 | 2525.1<br>2530.6<br>2541.7 | 0.110−5.02<br>0.110−5.01<br>0.110−4.98 | 1.0 | 1:3:5 | 1:2.166: 4.1666 |
| Fe I | 1853 | 289.0 | 2473.6<br>2480.5<br>2491.4 | 0.087−5.10<br>0.087−5.09<br>0.087−5.06 | 2.0 | 1:1.66:2.33 | 1:1.4823: 2.2412 |
| Fe I | 1607 | 403.4 | 2462.9<br>2473.6<br>2488.9 | 0.052−5.09<br>0.052−5.06<br>0.052−5.03 | 3.0 | 1:1.4:1.8 | 1: 1.27119: 1.75706 |
| Fe I | 1571 | 267.0 | 2818.3<br>2824.1<br>2833.3 | 0.958−5.36<br>0.958−5.35<br>0.958−5.33 | 3.0 | 1:1.4:1.8 | 1: 1.27119: 1.75706 |
| Fe I | 1445 | 216.8 | 2796.4<br>2805.3<br>2814.1 | 0.914−5.35<br>0.914−5.33<br>0.914−5.32 | 4.0 | 1:1.2857: 1.5714 | 1: 1.17608: 1.544020 |
| Fe I | 1204 | 240.7 | 2770.5<br>2779.0<br>2788.9 | 0.086−5.33<br>0.086−5.32<br>0.086−5.30 | 5.0 | 1:1.22:1.44 | 1: 1.122187: 1.424437 |
| Fe I | 6934 | 108.2 | 2724.4<br>2743.2<br>2772.9 | 0.087−4.64<br>0.087−4.60<br>0.087−4.56 | 2.0 | 1:1.66:2.33 | 1:1.4823: 2.2412 |
| Fe I | 5265 | 132.1 | 2511.6<br>2528.2<br>2550.4 | 0.052−4.99<br>0.052−4.96<br>0.052−4.91 | 3.0 | 1:1.4:1.8 | 1: 1.27119:<br>1.75706 |
| Fe I | 4615 | 115.8 | 2929.9<br>2948.7<br>2974.1 | 0.052−4.28<br>0.052−4.26<br>0.052−4.22 | 3.0 | 1:1.4:1.8 | 1: 1.27119:<br>1.75706 |
| Mn I | ∞ | 135.4 | 2795.6<br>2799.1<br>2801.9 | 0.000−4.43<br>0.000−4.43<br>0.000−4.42 | 2.5 | 1:1.5:2 | 1:1.3523:<br>1.94097 |
| Mn II | ∞ | 204.3 | 2576.9<br>2594.5<br>2606.5 | 0.000−4.81<br>0.000−4.78<br>0.000−4.76 | 3.0 | 1:1.4:1.8 | 1: 1.27119:<br>1.75706 |
| Ni II | 5228 | 149.2 | 2188.7<br>2225.6<br>2270.9 | 1.157−6.82<br>1.157−6.73<br>1.157−6.62 | 3.5 | 1:1.33:1.66 | 1: 1.21601:<br>1.6331 |
| Ni II | 4495 | 130.1 | 2159.4<br>2175.8<br>2207.4 | 1.254−7.00<br>1.254−6.95<br>1.254−6.87 | 2.5 | 1:1.5:2 | 1:1.3523:<br>1.94097 |
| KELT-20 | | | | | | | |
| C I | 3094 | 245.4 | 1656.3<br>1657.4<br>1657.9 | 0.002−7.49<br>0.002−7.48<br>0.002−7.48 | 1.0 | 1:3:5 | 1:2.166: 4.1666 |
| Fe I | ∞ | 335.3 | 2913.0<br>2937.8<br>2967.8 | 0.000−4.26<br>0.000−4.22<br>0.000−4.18 | 4.0 | 1:1.2857:<br>1.5714 | 1: 1.17608:<br>1.544020 |
| Fe I | ∞ | 929.1 | 2448.5<br>2463.4<br>2484.0 | 0.000−5.06<br>0.000−5.03<br>0.000−4.99 | 4.0 | 1:1.2857:<br>1.5714 | 1: 1.17608:<br>1.544020 |
| Fe I | 4231 | 660.1 | 2473.6<br>2480.5<br>2491.4 | 0.087−5.10<br>0.087−5.09<br>0.087−5.06 | 2.0 | 1:1.66:2.33 | 1:1.4823: 2.2412 |
| Fe I | 3826 | 226.5 | 2525.1<br>2530.6<br>2541.7 | 0.110−5.02<br>0.110−5.01<br>0.110−4.99 | 1.0 | 1:3:5 | 1:2.166: 4.1666 |
| Fe I | 3225 | 809.5 | 2462.9<br>2473.6<br>2488.9 | 0.052−5.09<br>0.052−5.06<br>0.052−5.03 | 3.0 | 1:1.4:1.8 | 1: 1.27119:<br>1.75706 |
| Fe I | 1109 | 173.0 | 2518.9<br>2529.9<br>2546.7 | 0.087−5.01<br>0.087−4.99<br>0.087−4.96 | 2.0 | 1:1.66:2.33 | 1:1.4823: 2.2412 |
| Fe I | 9351 | 187.0 | 2770.5<br>2779.0<br>2788.9 | 0.086−5.33<br>0.086−5.32<br>0.086−5.30 | 5.0 | 1:1.22:1.44 | 1: 1.122187:<br>1.424437 |
| Fe I | 8822 | 221.4 | 2511.6<br>2528.2<br>2550.4 | 0.052−4.99<br>0.052−4.96<br>0.052−4.91 | 3.0 | 1:1.4:1.8 | 1: 1.27119:<br>1.75706 |
| Fe I | 7159 | 107.4 | 2796.4<br>2805.3<br>2814.1 | 0.915−5.35<br>0.915−5.33<br>0.915−5.32 | 4.0 | 1:1.2857:<br>1.5714 | 1: 1.17608:<br>1.544020 |
| Fe I | 4537 | 771.4 | 2818.3<br>2824.1<br>2833.3 | 0.958−5.36<br>0.958−5.35<br>0.958−5.33 | 3.0 | 1:1.4:1.8 | 1: 1.27119:<br>1.75706 |
| Fe I | 4182 | 652.4 | 2724.4<br>2743.2<br>2772.9 | 0.087−4.64<br>0.087−4.61<br>0.087−4.56 | 2.0 | 1:1.66:2.33 | 1:1.4823: 2.2412 |
| Fe I | 4048 | 809.7 | 2570.4<br>2577.5<br>2585.3 | 0.858−5.68<br>0.858−5.67<br>0.858−5.65 | 5.0 | 1:1.22:1.44 | 1: 1.122187:<br>1.424437 |
| Fe I | 3430 | 514.4 | 2594.9<br>2600.3<br>2607.6 | 0.914−5.69<br>0.914−5.68<br>0.914−5.67 | 4.0 | 1:1.2857:<br>1.5714 | 1: 1.17608:<br>1.544020 |
| Fe I | 2570 | 436.9 | 2707.4<br>2744.4<br>2755.2 | 0.958−5.54<br>0.958−5.48<br>0.958−5.46 | 3.0 | 1:1.4:1.8 | 1: 1.27119:<br>1.75706 |
| Fe I | 2564 | 435.9 | 2615.3<br>2618.8<br>2624.3 | 0.958−5.70<br>0.958−5.69<br>0.958−5.68 | 3.0 | 1:1.4:1.8 | 1: 1.27119:<br>1.75706 |
| Fe I | 2339 | 350.8 | 2781.7<br>2798.6<br>2807.8 | 0.915−5.37<br>0.915−5.34<br>0.915−5.33 | 4.0 | 1:1.2857:<br>1.5714 | 1: 1.17608:<br>1.544020 |
| Fe I | 1849 | 277.4 | 2690.0<br>2699.9<br>2712.5 | 0.915−5.52<br>0.915−5.51<br>0.915−5.49 | 4.0 | 1:1.2857:<br>1.5714 | 1: 1.17608:<br>1.544020 |
| Fe I | 1841 | 497.0 | 2719.2<br>2726.4<br>2763.9 | 0.990−5.55<br>0.990−5.54<br>0.990−5.48 | 2.0 | 1:1.66:2.33 | 1:1.4823: 2.2412 |
| Fe I | 1497 | 254.4 | 2743.1<br>2762.8<br>2795.5 | 0.958−5.48<br>0.958−5.45<br>0.958−5.39 | 3.0 | 1:1.4:1.8 | 1: 1.27119:<br>1.75706 |
| Fe I | 1379 | 469.0 | 2726.9<br>2731.8<br>2739.0 | 1.011−5.56<br>1.011−5.55<br>1.011−5.54 | 1.0 | 1:3:5 | 1:2.166: 4.1666 |
| Fe I | 1353 | 202.9 | 2480.4<br>2486.7<br>2497.3 | 0.915−5.91<br>0.915−5.90<br>0.915−5.88 | 4.0 | 1:1.2857:<br>1.5714 | 1: 1.17608:<br>1.544020 |
| Fe I | 1349 | 798.8 | 2295.1<br>2297.6<br>2309.7 | 0.110−5.51<br>0.110−5.51<br>0.110−5.48 | 1.0 | 1:3:5 | 1:2.166: 4.1666 |
| Fe I | 1219 | 190.1 | 2942.2<br>2954.8<br>2974.0 | 0.087−4.30<br>0.087−4.28<br>0.087−4.26 | 2.0 | 1:1.66:2.33 | 1:1.4823: 2.2412 |
| Fe I | 1154 | 230.9 | 2459.3<br>2469.6<br>2473.1 | 0.859−5.90<br>0.859−5.88<br>0.859−5.87 | 5.0 | 1:1.22:1.44 | 1: 1.122187:<br>1.424437 |
| Fe I | 1114 | 378.7 | 2754.5<br>2758.1<br>2775.5 | 1.011−5.51<br>1.011−5.51<br>1.011−5.48 | 1.0 | 1:3:5 | 1:2.166: 4.1666 |
| Fe I | 9369 | 235.2 | 2929.9<br>2948.7<br>2974.1 | 0.052−4.28<br>0.052−4.26<br>0.052−4.22 | 3.0 | 1:1.4:1.8 | 1: 1.27119:<br>1.75706 |
| Fe I | 7035 | 190.0 | 2745.3<br>2762.6<br>2782.7 | 0.990−5.51<br>0.990−5.48<br>0.990−5.45 | 2.0 | 1:1.66:2.33 | 1:1.4823: 2.2412 |
| Fe I | 4070 | 634.9 | 2288.0<br>2299.9<br>2313.8 | 0.087−5.51<br>0.087−5.48<br>0.087−5.45 | 2.0 | 1:1.66:2.33 | 1:1.4823: 2.2412 |
| Fe I | 3876 | 104.6 | 2734.8<br>2747.8<br>2754.8 | 0.990−5.52<br>0.990−5.50<br>0.990−5.49 | 2.0 | 1:1.66:2.33 | 1:1.4823: 2.2412 |
| Fe I | 2289 | 574.6 | 2284.8<br>2298.5<br>2321.1 | 0.052−5.48<br>0.052−5.45<br>0.052−5.39 | 3.0 | 1:1.4:1.8 | 1: 1.27119: 1.75706 |
| Fe I | 2006 | 341.0 | 2715.7<br>2725.8<br>2735.4 | 0.958−5.52<br>0.958−5.51<br>0.958−5.49 | 3.0 | 1:1.4:1.8 | 1: 1.27119: 1.75706 |
| Fe I | 5881 | 158.8 | 2849.6<br>2867.5<br>2894.6 | 0.990−5.34<br>0.990−5.31<br>0.990−5.27 | 2.0 | 1:1.66:2.33 | 1:1.4823: 2.2412 |
| Fe II | 3832 | 276.3 | 2405.2<br>2407.4<br>2411.3 | 0.107−5.26<br>0.107−5.26<br>0.107−5.25 | 1.5 | 1:2:3 | 1: 1.727: 2.8187 |
| Fe II | 3539 | 255.1 | 2614.6<br>2621.2<br>2631.8 | 0.107−4.85<br>0.107−4.84<br>0.107−4.82 | 1.5 | 1:2:3 | 1: 1.727: 2.8187 |
| Fe II | 1879 | 296.9 | 2396.1<br>2400.0<br>2405.6 | 0.083−5.26<br>0.083−5.25<br>0.083−5.24 | 2.5 | 1:1.5:2 | 1:1.3523: 1.94097 |
| Fe II | 1859 | 399.8 | 2383.8<br>2389.4<br>2396.4 | 0.048−5.25<br>0.048−5.24<br>0.048−5.22 | 3.5 | 1:1.33:1.66 | 1: 1.21601: 1.6331 |
| Fe II | 1317 | 208.1 | 2607.9<br>2618.4<br>2632.1 | 0.083−4.84<br>0.083−4.82<br>0.083−4.79 | 2.5 | 1:1.5:2 | 1:1.3523: 1.94097 |
| Fe II | 1055 | 226.9 | 2599.1<br>2612.7<br>2626.5 | 0.048−4.82<br>0.048−4.79<br>0.048−4.77 | 3.5 | 1:1.33:1.66 | 1: 1.21601: 1.6331 |
| Fe II | 4589 | 561.2 | 2710.2<br>2725.7<br>2750.1 | 1.040−5.62<br>1.040−5.89<br>1.040−5.55 | 2.5 | 1:1.5:2 | 1:1.3523: 1.94097 |
| Fe II | 4567 | 721.6 | 2328.1<br>2349.0<br>2381.5 | 0.083−5.41<br>0.083−5.36<br>0.083−5.29 | 2.5 | 1:1.5:2 | 1:1.3523: 1.94097 |
| Fe II | 4003 | 562.2 | 2693.6<br>2717.5<br>2756.6 | 0.986−5.59<br>0.986−5.55<br>0.986−5.48 | 3.5 | 1:1.33:1.66 | 1: 1.21601: 1.6331 |
| Fe II | 2071 | 253.2 | 2728.3<br>2747.8<br>2773.5 | 1.040−5.58<br>1.040−5.55<br>1.040−5.51 | 2.5 | 1:1.5:2 | 1:1.3523: 1.94097 |
| Fe II | 2005 | 283.8 | 1608.5<br>1610.9<br>1612.8 | 0.232−7.94<br>0.232−7.93<br>0.232−7.92 | 4.5 | 1:1.25:1.5 | 1: 1.1458574: 1.476857 |
| Fe II | 1299 | 163.7 | 2737.8<br>2750.0<br>2769.8 | 1.076−5.60<br>1.076−5.58<br>1.076−5.55 | 1.5 | 1:2:3 | 1: 1.727: 2.8187 |
| Fe II | 7.602 | 958.0 | 2567.7<br>2583.4<br>2611.9 | 1.076−5.90<br>1.076−5.88<br>1.076−5.82 | 1.5 | 1:2:3 | 1: 1.727: 2.8187 |
| Fe II | 4423 | 698.8 | 1625.9<br>1629.2<br>1631.1 | 0.083−7.71<br>0.083−7.69<br>0.083−7.68 | 2.5 | 1:1.5:2 | 1:1.3523: 1.94097 |
| Fe II | 3405 | 137.4 | 2369.3<br>2384.0<br>2403.3 | 0.352−5.58<br>0.352−5.55<br>0.352−5.51 | 2.5 | 1:1.5:2 | 1:1.3523: 1.94097 |
| Fe II | 1298 | 617.9 | 2479.2 | 2.635−7.64 | 6.5 | 1:1.16:1.33 | 1: 1.0744337: |
| | | | 2482.9 | 2.635−7.63 | | | 1.3191436 |
| | | | 2494.0 | 2.635−7.61 | | | |
| Fe II | 9213 | 116.1 | 1646.2 | 1.076−8.61 | 1.5 | 1:2:3 | 1: 1.727: 2.8187 |
| | | | 1649.6 | 1.076−8.59 | | | |
| | | | 1655.0 | 1.076−8.57 | | | |
| Fe II | 8875 | 358.1 | 1568.0 | 0.352−8.26 | 2.5 | 1:1.5:2 | 1:1.3523: |
| | | | 1570.2 | 0.352−8.25 | | | 1.94097 |
| | | | 1573.8 | 0.352−8.23 | | | |
| Fe II | 7017 | 420.6 | 2344.7 | 0.301−5.59 | 3.5 | 1:1.33:1.66 | 1: 1.21601: |
| | | | 2362.7 | 0.301−5.55 | | | 1.6331 |
| | | | 2392.2 | 0.301−5.48 | | | |
| Fe II | 6845 | 276.2 | 2355.6 | 0.352−5.62 | 2.5 | 1:1.5:2 | 1:1.3523: |
| | | | 2367.3 | 0.352−5.59 | | | 1.94097 |
| | | | 2385.7 | 0.352−5.55 | | | |
| Fe II | 6607 | 294.7 | 2490.2 | 2.657−7.64 | 5.5 | 1:1.2:1.4 | 1: 1.1031523: |
| | | | 2493.9 | 2.657−7.63 | | | 1.382391 |
| | | | 2494.9 | 2.657−7.63 | | | |
| Fe II | 6204 | 276.7 | 2527.9 | 2.657−7.56 | 5.5 | 1:1.2:1.4 | 1: 1.1031523: |
| | | | 2534.4 | 2.657−7.55 | | | 1.382391 |
| | | | 2537.6 | 2.657−7.54 | | | |
| Fe II | 5173 | 222.0 | 2526.8 | 2.676−7.58 | 4.5 | 1:1.25:1.5 | 1: 1.1458574: |
| | | | 2537.6 | 2.676−7.56 | | | 1.476857 |
| | | | 2544.1 | 2.676−7.55 | | | |
| Fe II | 3186 | 455.2 | 2530.3 | 2.704−7.60 | 1.5 | 1:2:3 | 1: 1.727: 2.8187 |
| | | | 2549.5 | 2.704−7.57 | | | |
| | | | 2591.3 | 2.704−7.49 | | | |
| Fe II | 3058 | 354.4 | 2373.5 | 2.583−7.81 | 2.5 | 1:1.5:2 | 1:1.3523: |
| | | | 2389.0 | 2.583−7.77 | | | 1.94097 |
| | | | 2445.3 | 2.583−7.65 | | | |
| Fe II | 2684 | 161.3 | 2467.4 | 2.856−7.88 | 1.5 | 1:2:3 | 1: 1.727: 2.8187 |
| | | | 2473.2 | 2.856−7.87 | | | |
| | | | 2485.0 | 2.856−7.84 | | | |
| Fe II | 2552 | 480.5 | 2429.1 | 3.903−9.01 | 3.5 | 1:1.33:1.66 | 1: 1.21601: |
| | | | 2446.9 | 3.903−8.97 | | | 1.6331 |
| | | | 2453.7 | 3.903−8.96 | | | |
| Fe II | 2027 | 311.2 | 2447.0 | 3.199−8.27 | 4.5 | 1:1.25:1.5 | 1: 1.1458574: |
| | | | 2452.0 | 3.199−8.26 | | | 1.476857 |
| | | | 2459.5 | 3.199−8.24 | | | |
| Fe II | 1943 | 277.6 | 2430.1 | 2.704−7.81 | 1.5 | 1:2:3 | 1: 1.727: 2.8187 |
| | | | 2438.0 | 2.704−7.79 | | | |
| | | | 2446.3 | 2.704−7.77 | | | |
| Fe II | 1915 | 832.8 | 1785.3 | 2.891−9.84 | 2.5 | 1:1.5:2 | 1:1.3523: |
| | | | 1786.8 | 2.891−9.83 | | | 1.94097 |
| | | | 1788.0 | 2.891−9.83 | | | |
| Fe II | 1847 | 222.6 | 2415.2 | 2.807−7.94 | 4.5 | 1:1.25:1.5 | 1: 1.1458574: |
| | | | 2420.6 | 2.807−7.93 | | | 1.476857 |
| | | | 2424.9 | 2.807−7.92 | | | |
| Fe II | 1423 | 106.8 | 2544.2 | 2.844−7.72 | 2.5 | 1:1.5:2 | 1:1.3523: |
| | | | 2550.2 | 2.844−7.71 | | | 1.94097 |
| | | | 2555.8 | 2.844−7.70 | | | |
| Fe II | 1090 | 124.0 | 2465.7 | 3.230−8.26 | 2.5 | 1:1.5:2 | 1:1.3523: |
| | | | 2471.2 | 3.230−8.25 | | | 1.94097 |
| | | | 2480.0 | 3.230−8.23 | | | |
| Fe II | 1060 | 122.3 | 2494.0<br>2503.1<br>2518.0 | 3.221−8.19<br>3.221−8.17<br>3.221−8.15 | 3.5 | 1:1.33:1.66 | 1: 1.21601: 1.6331 |
| Fe II | 8936 | 823.6 | 2541.4<br>2541.9<br>2547.4 | 2.828−7.71<br>2.828−7.71<br>2.828−7.70 | 3.5 | 1:1.33:1.66 | 1: 1.21601: 1.6331 |
| Fe II | 8717 | 387.0 | 2359.9<br>2362.4<br>2367.6 | 2.692−7.95<br>2.692−7.94<br>2.692−7.93 | 3.5 | 1:1.33:1.66 | 1: 1.21601: 1.6331 |
| Fe II | 8573 | 989.6 | 2466.7<br>2475.5<br>2498.1 | 3.221−8.25<br>3.221−8.23<br>3.221−8.18 | 3.5 | 1:1.33:1.66 | 1: 1.21601: 1.6331 |
| Fe II | 7197 | 319.5 | 2542.6<br>2548.1<br>2557.8 | 2.692−7.57<br>2.692−7.56<br>2.692−7.54 | 3.5 | 1:1.33:1.66 | 1: 1.21601: 1.6331 |
| Fe II | 5436 | 834.6 | 2492.1<br>2506.8<br>2414.2 | 3.199−8.17<br>3.199−8.15<br>3.199−8.13 | 4.5 | 1:1.25:1.5 | 1: 1.1458574: 1.476857 |
| Fe II | 4867 | 220.7 | 2004.6<br>2008.4<br>2016.1 | 1.695−7.88<br>1.695−7.87<br>1.695−7.84 | 1.5 | 1:2:3 | 1: 1.727: 2.8187 |
| Fe II | 4621 | 209.5 | 1696.5<br>1699.2<br>1701.9 | 1.695−9.00<br>1.695−8.99<br>1.695−8.98 | 1.5 | 1:2:3 | 1: 1.727: 2.8187 |
| Fe II | 4599 | 197.4 | 2355.2<br>2360.3<br>2364.4 | 2.676−7.94<br>2.676−7.93<br>2.676−7.92 | 4.5 | 1:1.25:1.5 | 1: 1.1458574: 1.476857 |
| Fe II | 4277 | 394.2 | 2615.7<br>2621.5<br>2631.8 | 2.828−7.57<br>2.828−7.56<br>2.828−7.54 | 3.5 | 1:1.33:1.66 | 1: 1.21601: 1.6331 |
| Fe II | 2774 | 334.3 | 2609.6<br>2619.9<br>2632.4 | 2.807−7.56<br>2.807−7.54<br>2.807−7.52 | 4.5 | 1:1.25:1.5 | 1: 1.1458574: 1.476857 |
| Fe II | 2151 | 107.7 | 1689.8<br>1690.8<br>1693.5 | 1.671−9.01<br>1.671−9.00<br>1.671−8.99 | 2.5 | 1:1.5:2 | 1:1.3523: 1.94097 |
| Mn I | ∞ | 237.5 | 2795.6<br>2799.1<br>2801.9 | 0.000−4.43<br>0.000−4.43<br>0.000−4.42 | 2.5 | 1:1.5:2 | 1:1.3523: 1.94097 |
| Mn II | ∞ | 165.2 | 2576.9<br>2594.5<br>2606.5 | 0.000−4.81<br>0.000−4.78<br>0.000−4.76 | 3.0 | 1:1.4:1.8 | 1: 1.27119: 1.75706 |
| Ni II | 7239 | 209.5 | 2159.4<br>2175.8<br>2207.4 | 1.254−7.00<br>1.254−6.95<br>1.254−6.87 | 2.5 | 1:1.5:2 | 1:1.3523: 1.94097 |
| Ni II | 6954 | 198.5 | 2188.7<br>2225.6<br>2270.9 | 1.157−6.82<br>1.157−6.73<br>1.157−6.62 | 3.5 | 1:1.33:1.66 | 1: 1.21601: 1.6331 |
| Ni II | 2484 | 709.0 | 2139.3<br>2169.8<br>2211.1 | 1.157−6,95<br>1.157−6.87<br>1.157−6.76 | 3.5 | 1:1.33:1.66 | 1: 1.21601: 1.6331 |
| Sc II | ∞ | 204.0 | 2541.6<br>2556.6<br>2564.0 | 0.000−4.88<br>0.000−4.85<br>0.000−4.84 | 1.0 | 1:3:5 | 1:2.166: 4.1666 |